# On Extrinsic Information of Good Codes Operating Over Discrete Memoryless Channels

Michael Peleg, Amichai Sanderovich and Shlomo Shamai (Shitz)

*Abstract:* We show that the Extrinsic Information about the coded bits of any good (capacity achieving) code operating over a wide class of discrete memoryless channels (DMC) is zero when channel capacity is below the code rate and positive constant otherwise, that is, the Extrinsic Information Transfer (EXIT) chart is a step function of channel quality, for any capacity achieving code. It follows that, for a common class of iterative receivers where the error correcting decoder must operate at first iteration at rate above capacity (such as in turbo equalization, turbo channel estimation, parallel and serial concatenated coding and the like), classical good codes which achieve capacity over the DMC are not effective and should be replaced by different new ones. Another meaning of the results is that a good code operating at rate above channel capacity falls apart into its individual transmitted symbols in the sense that all the information about a coded transmitted symbol is contained in the corresponding received symbol and no information about it can be inferred from the other received symbols.

This report comprises two stand-alone parts. Part 1 treats the binary input additive white Gaussian noise channel. Part 2 extends the results to the symmetric binary channel and to the binary erasure channel and provides an heuristic extension to wider class of channel models such as common fading models combined with QPSK and other inputs which might be rigorized by further work.

The authors are with the Technion-Israel Institute of Technology, Technion city, Haifa, Israel

# Part 1

# On Extrinsic Information of Good Binary Codes Operating Over Gaussian Channels

M. Peleg, A. Sanderovich and S. Shamai (Shitz)

*Abstract:* We show that the Extrinsic Information about the coded bits of any good (capacity achieving) binary code operating over a Gaussian channel is zero when the channel capacity is lower then code rate and unity when capacity exceeds the code rate, that is, the Extrinsic Information Transfer (EXIT) chart is a step function of the signal to noise ratio and independent of the code. It follows that, for a common class of iterative receivers where the error correcting decoder must operate at first iteration at rate above capacity (such as in turbo equalization, iterative channel estimation, parallel and serial concatenated coding and the like), classical good codes which achieve capacity over the Additive White Gaussian Noise Channel are not effective and should be replaced by different new ones.

*I Introduction*

In this letter we derive the Extrinsic Information Transfer (EXIT) chart of asymptotically long binary codes which achieve a vanishing probability of error over the Additive White Gaussian Noise (AWGN) channel at code rates below the channel capacity. We denote such codes as "good codes" in the following. The results provide an insight about corresponding iterative receivers designed to approach the channel capacity assuming asymptotically long codewords.

It is well known that when a good Error Correcting Code (ECC) is used to transmit information over a channel the capacity of which is lower than the code rate then the error rate is high. This scenario actually occurs at the first decoding iteration performed by the new iterative receivers based on the turbo principle, see fig.1, where some preprocessor such as equalizer, multi-user detector [1], phase estimator or other precedes the decoder and employs the decoder outputs to improve the channel presented to the decoder over the successive iterations. The ECC code may be any code, including an iteratively decodable turbo or LDPC code.

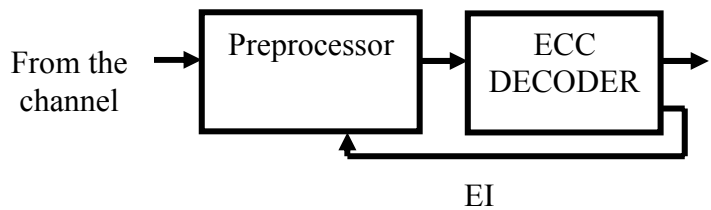

*Figure 1. Iterative receiver*

If the whole iterative receiver is designed to approach capacity and if the iterative feedback to the preprocessor is really required then at the first iteration the ECC decoder is presented with a channel the capacity of which is below the code rate while at the next iterations the preprocessor will improve the ECC decoder input and enable errorless decoding. To achieve this, the ECC decoder must pass some useful information to the preprocessor at the first iteration, while operating over channel the capacity of which is below the code rate. The relevant information to be passed is the well known Extrinsic Information (EI) to be defined below.

In fact, serially concatenated turbo codes can also be represented by the structure of fig.1 where the preprocessor is the decoder of the inner component code. Parallel concatenated turbo codes are decoded by a similar structure and the operation at rate above channel capacity at first iteration is then also clearly required since the component code is presented only with a subset of the channel output symbols.

In the following derivation we will show formally that over AWGN channels, codes considered to be good in the classical sense, provide no EI at all in this setting and thus eliminate any improvement by the iterative feedback in fig. 1. So the search for different codes fitting the new iterative systems such as performed in [1], [2],[3] and others is indeed essential if the iterative receiver is to perform better than, say, separate equalization and decoding. Also, classical good codes cannot perform well as outer codes in a serially concatenated turbo code or component codes in parallel concatenated turbo codes. This was indicated first in [4] where increasing the constraint length of a convolutional component code rendered the iterative feedback of a turbo decoder ineffective. More precisely, we will show that the Extrinsic Information Transfer (EXIT) chart of any good binary code over a memoryless AWGN channel is a step function, the EI being zero for Signal to Noise Ratio (SNR) at which the channel capacity is below the code rate and unity at SNR larger than this.

Notations: Mutual information is denoted by I, entropy by H and statistical expectation by E. Probability and probability density function are denoted by P and p respectively and bold letters denote vectors.

## II Models and definitions

1. *Channel model*:

We examine a Binary Input Additive White Gaussian Noise channel (BI-AWGN channel

$$y = \sqrt{s}(2x-1) + n \qquad (1)$$

where (2x-1) is the transmitted signal, -1 or 1, x is the corresponding bit at channel input, 0 or 1 and n is a Gaussian random variable with zero mean and unit variance. The signal power, denoted s, is equal to the SNR. The channel is characterized by the Gaussian probability p(y|x) and C denotes the channel capacity.

For a pair of SNR values $s^l < s^h$, the channel characterized by $s=s^l$, , can be described as physically degraded with respect to a channel characterized by $s^h$, where the superscripts h and l denote 'higher' and 'lower' SNR respectively. The outputs of the two channels are denoted $y^l$ and $y^h$. By physically degraded we mean that $x_i$, $y^l$ and $y^h$ form following Markov chain

$$x_i - y_i^h - y_i^l \qquad (2)$$

arising from the possibility to obtain $y^l$ from $y^h$ by adding independent Gaussian noise and scaling to conform to (1).

*2. The good code:*

We desire to transmit information **U**. We do so in the standard manner of transmitting a codeword **x**, a vector of n channel symbols $x_i$ belonging to an asymptotically long good code X of rate R. The selection of the transmitted codewords is determined by **U** and is equi-probable. The received vector is denoted **y**. The code is such that it achieves vanishing error probability at channel SNR $s=s^0$ for which $R>C(s^0)-\varepsilon'$ where $\varepsilon'$ is positive and can be made as small as desired by increasing N. A well known result is then

$$\frac{1}{n}I[\mathbf{x};\mathbf{y}(s^0)] \geq C(s^0)-\varepsilon' \qquad (3)$$

*3. The Extrinsic Information:*

We are interested in a symbol $x_i$, which is a symbol at the i'th position in **x**. We define $\mathbf{x'}_i$ as **x** with $x_i$ excluded and correspondingly $\mathbf{y'}_i$ as **y** with $y_i$ excluded.

We denote $z_i$ the complete information obtainable from $\mathbf{y'}_i$ about $x_i$, known as the extrinsic information, see for example [4]. The extrinsic information $z_i$ can be expressed for example as the Logarithmic Likelihood Ratio (LLR) of $x_i$ or as $P(x_i=1|\mathbf{y'}_i)$.

The average extrinsic information measure is then defined as

$$EI = \frac{1}{n}\sum_{i=1}^{n} I(x_i;\mathbf{y'}_i) = \frac{1}{n}\sum_{i=1}^{n} I(x_i;z_i) \qquad (4)$$

When $x_i$ is given, then $y_i$ is independent of $\mathbf{y'}_i$, that is $p(y_i|x_i,\mathbf{y'}_i) = p(y_i|x_i)$. This extends the Markov chain (2) to

$$\mathbf{y'}_i^l - \mathbf{y'}_i^h - x_i - y_i^h - y_i^l \tag{5}$$

$$z_i - x_i - y_i^h - y_i^l \tag{6}$$

Furthermore, due to this Markov chain and the data processing theorem we have $I(\mathbf{y'}_i^h; x_i) \geq I(\mathbf{y'}_i^l; x_i)$, thus EI is a non-decreasing function of s.

$$EI(s^h) \geq EI(s^l) \tag{7}$$

*III EXIT chart of good codes*

When the capacity C is strictly above the code rate R, we have perfect decoding for asymptotically long good codes, even if the single symbol $y_i$ is removed (erased) before the decoding. Thus we have for any small positive $\varepsilon'$ and large n:

$$R < C - \varepsilon' \rightarrow EI = 1 \tag{8}$$

This intuitive attribute of good codes is verified in Appendix A for finite s.

The central result of this letter is the following proposition and its method of proof:

**_Proposition_**: *The average EI, eq. (4), about the coded bits $x_i$ of a good binary code operating over an BI-AWGN channel, the capacity of which is lower then the code rate, is zero. (More preciously smaller then any positive $\varepsilon_2$ for n large enough.)*

Proof: It is well known that, when the code rate R is at or above capacity, good codes mimic closely the channel output statistics of a capacity achieving identically and independently distributed ( i.i.d.) input [5, Theorem 15].

Specifically, we prove in Appendix B that for all $s<s^0$, the mutual information I (**x**;**y**) over the channel with the good code is similar to the symbol wise mutual information. That is for any small $\varepsilon' > 0$ and sufficiently large n:

$$0 \leq \frac{1}{n}\sum_{i=1}^{n} I[x_i; y_i(s)] - \frac{1}{n} I[\mathbf{x}; \mathbf{y}(s)] = \gamma(s) \leq \varepsilon' = \varepsilon^3 \tag{9}$$

where $\gamma$ is a non-decreasing function of s. The substitution $\varepsilon' = \varepsilon^3$ will be required below.

We shall need to upper-bound the derivative of (9) with respect to s:

$$ID(s) \triangleq \frac{d}{ds}\{\frac{1}{n}\sum_{i=1}^{n} I[x_i; y_i(s)] - \frac{1}{n}I[\mathbf{x};\mathbf{y}(s)]\} = \frac{d}{ds}\gamma(s) \qquad (10)$$

Since $\gamma$ is non-decreasing, ID is non-negative and its average over an interval of $s^0 - \Delta$ to $s^0$ cannot exceed $\varepsilon^3/\Delta$, otherwise its integral (9) would exceed $\varepsilon^3$. We shall choose $\Delta = \varepsilon$ to limit the average ID to $\varepsilon^2$. Thus there is some $s=s_t$ in the above interval, $\varepsilon$ within $s^0$, for which the absolute value of ID is bounded by

$$ID(s_t) \leq \varepsilon^2 \qquad (11)$$

Due to (7), vanishing EI at $s_t$ implies vanishing EI at all smaller values of s, so it is sufficient to prove vanishing EI at $s_t$.

In [6] the Minimum Mean Square Error (MMSE), when estimating a general input of a Gaussian channel using the channel output, is linked to the derivative with respect to the SNR of the relevant mutual information. Using [6, theorem 2] while the transmitted signal **Hx** in [6] is our (2**x**-1), see (1), and the constant 1 does not influence the estimation errors, we can see that the sum of the estimation errors of all the symbols $x_i$ is:

$$\sum_{i=1}^{n} MMSE(x_i | \mathbf{y}) = \frac{d}{ds} 0.5 I[\mathbf{x};\mathbf{y}(s)] \quad , \qquad (12)$$

where MMSE($x_i|\mathbf{y}$) denotes the MMSE of an individual bit $x_i$ obtained by optimally estimating $x_i$ from **y**. The factor before the sum in (12) is 0.5 rather then 2 in [6] because the transmitted signal (1) is 2x-1 rather then x. However, for any $x_i$, a similar estimation error can be achieved using merely the single received symbol $y_i$, see [6, eq. (1)]:

$$\sum_{i=1}^{n} \text{MMSE}(x_i | y_i) = \frac{d}{ds} 0.5 \sum_{i=1}^{n} I[x_i; y_i(s)] \qquad (13)$$

By (10) to (13) we have then

$$0 \leq \frac{1}{n}\sum_{i=1}^{n} \text{MMSE}(x_i | y_i) - \frac{1}{n}\sum_{i=1}^{n} \text{MMSE}(x_i | \mathbf{y}) \leq 0.5\varepsilon^2$$

$$0 \leq \frac{1}{n}\sum_{i=1}^{n} [\text{MMSE}(x_i | y_i) - \text{MMSE}(x_i | \mathbf{y})] \leq 0.5\varepsilon^2 \qquad (14)$$

Clearly each element of the above sum is positive and their average is upper bounded by $0.5\varepsilon^2$. This implies that each element is upper bounded by $\varepsilon$, except at most $0.5n\varepsilon$ elements which may be larger (with $0.5n\varepsilon$ elements larger than $\varepsilon$, (14) will be violated). This vanishing proportion of elements can contribute only $0.5\varepsilon$ bits to the average EI (4) because the EI for each bit is bounded by 1, so we can disregard them in our proof of vanishing average EI, eq. (4), and use

$$\text{MMSE}(x_i | y_i) - \text{MMSE}(x_i | \mathbf{y}) \leq \varepsilon \qquad (15)$$

Thus, at $s=s_t$, the MMSE estimation error of xi using **y** is nearly the same as if only $y_i$ was used.

The MMSE estimate of $x_i$, valued 0 or 1, is its conditional expectation

$$\hat{x}_i = 0 * P(x_i=0|\mathbf{y'}_i, y_i) + 1 * P(x_i=1|\mathbf{y'}_i, y_i)$$

$$\hat{x}_i(\mathbf{y'}_i, y_i) = P(x_i=1|\mathbf{y'}_i, y_i). \qquad (16)$$

and $\hat{x}_i(y_i) = P(x_i=1|y_i)$.
Eq. (15), (16) imply that $y_i$ is an approximate sufficient statistics, where the full statistics is $\mathbf{y}=(y_i, \mathbf{y}_i')$. This immediately implies (17) due to continuity.
as shown in more detail in appendix C:

$$E[P(x_i=1|\mathbf{y'}_i, y_i) - P(x_i=1|y_i)]^2 \leq \varepsilon \qquad (17)$$

Thus $P(x_i=1|\mathbf{y'}_i, y_i) \cong P(x_i=1|y_i)$, showing that $\mathbf{y'}_i$ cannot provide additional information about $x_i$ when $y_i$ is known.

To establish ( for any small positive $\varepsilon_2$ )

$$EI = \frac{1}{n}\sum_{i=1}^{n} I(x_i;\mathbf{y}'_i) \leq \varepsilon_2 \qquad (18)$$

we have to show that $\mathbf{y}'_i$ cannot provide information about $x_i$ also when $y_i$ is not known. This is equivalent to $P(x_i=1|\mathbf{y}'_i)=P(x_i=1)$. We present in the following the principles leading to (18), while a detailed but tedious proof is presented in appendix D.

Since $P(x_i=1|\mathbf{y}'_i,y_i)$ and $P(x_i=1|y_i)$ determine the Log Likelihood Ratios (LLR) which are closely coupled to mutual information, (17) leads to

$$I(x_i;\mathbf{y}'_i,y_i)-I(x_i;y_i)<\varepsilon 1 \qquad (19)$$

for some $\varepsilon 1$ proportional to $\varepsilon$. Next we shall use the bounds on information obtained by combining the outputs of independent channels [7]. The channel outputs $y_i$ and $\mathbf{y}'_i$ can be considered independent sources of information about $x_i$ by (5). Thus $y_i$ and $\mathbf{y}'_i$ are outputs of parallel broadcast channels in the sense of [7]. Also the channel (1) $x_i$ to $y_i$ is symmetric and $x_i$ is nearly uniformly distributed over -1 and 1 for all but a vanishing proportion of symbols $x_i$ (otherwise the code cannot achieve capacity, see [8] and references therein) and those can be disregarded since they cannot influence the average EI (4) because the contribution of one symbol to the average EI is limited to $1/n$.

The bounds [7] hold for x distributed uniformly according to $P(x=1) =0.5$, however since the bounds based on mutual information combining are all continuous functions of the parameter $P(x=1)$ describing the distribution of x, small deviations from $P(x=1) =0.5$ inflict small deviation on the outputs.

Under those conditions it follows from [7, theorem 2] that when $I(y_i ; x_i )$ and $I(\mathbf{y}' ; x_i)$ are given, then $I(\mathbf{y}'_i, y_i ; x_i )$, the information about $x_i$ obtained by combining both $y_i$ and $\mathbf{y}'_i$, is lower bounded by the one obtained when replacing $\mathbf{y}'_i$ by the output B of a Binary Symmetric Channel (BSC) transmitting $x_i$, with $I(B ; x_i) = I(\mathbf{y}'_i ; x_i)$. Straightforward calculation reveals that for the Gaussian channel with input $x_i$ and output $y_i$ and BSC with output B we have $I(B, y_i ; x_i) > I(y_i ; x_i) + \alpha I(B ; x_i)$ for some positive $\alpha$. Thus $I(\mathbf{y}'_i, y_i ; x_i) > I(y_i ; x_i) + \alpha I(\mathbf{y}'_i ; x_i)$. This together with (19) implies:

$$I(x_i ; \mathbf{y}'_i) \leq \varepsilon_2. \qquad (20)$$

for any small positive $\varepsilon_2$. This establishes the proposition (18); a detailed proof is available in appendix D. ∎

By the data processing theorem for the Markov chain (5) we have $I(x_i ; \mathbf{y}'_i) \geq I(y_i ; \mathbf{y}'_i)$, thus

$$\frac{1}{n} \sum_{i=1}^{n} I(y_i ; \mathbf{y}'_i) \leq \varepsilon_2 \qquad (21)$$

when channel capacity is below the code rate. This attribute cannot be derived directly from the results of [5] and does not hold when capacity exceeds slightly the code rate and successful decoding occurs.

Remarks: When capacity exceeds the code rate there is a sharp transition of the EI since $I(\mathbf{x},\mathbf{y})$ is not determined by (9) any more but reaches a plateau at the code rate and the MMSE, which is proportional to the derivative of $I(\mathbf{x},\mathbf{y})$ with respect to s=SNR by (12), goes to zero abruptly. This transition takes place over a small region of s for which the difference (9) is very small but, significantly, not zero. A similar transition of $\frac{1}{n} \sum_{i=1}^{n} I(y_i ; \mathbf{y}')$, see (21), occurs in the same region.

Clearly the area A under the EI versus $I(x_i;y_i)$ curve, $A = \int_0^1 EI \, d[I(x_i;y_i)]$, equals 1-R, thus the step function EXIT chart derived here for good binary code over the AWGN channel conforms to the "EXIT chart area property" of outer codes which was proved in [9] for any code over the binary erasure (BEC) channel . This property over the BEC channel together with (8), which is easy to verify also for the BEC channel, imply that (18) holds for good codes over the BEC channel too.

*Conclusions:*

The EXIT chart of any good binary code operating over a Gaussian channel is a step function, zero at channel capacities below the code rate and unity at capacities above the code rate. Thus codes good over the AWGN channels are very inefficient when used in an iterative receiver of the type presented in figure 1 which includes turbo-equalization, iterative multi-user receivers and serially concatenated codes as special cases. Interestingly, the step function EXIT chart derived here for the AWGN channel conforms to the EXIT chart area property derived in [9] for the erasure channel. Furthermore, good code operating at rate above channel capacity falls apart into its individual transmitted symbols in the sense that all the information about a coded bit $x_i$ is contained in the corresponding received symbol $y_i$ and no information about $x_i$ can be inferred from the other received symbols, neither alone, see (18) and neither as supplement to $y_i$, see (17).

It is of interest if the main result of this letter, namely (18), holds for more general memoryless channels. Based on [9], the results hold for the BEC channel as explained above, [10] outlines an extension to a wider class of M-ary input memoryless channels using the concept of GEXIT [11], and the arguments presented in [1] suggest extension to any memoryless channel for binary random codes.

## *Appendix A*

This appendix verifies (8) for finite s. When $R<C-\varepsilon'$, the symbol $x_i$ is decoded with zero error probability. Furthermore, by the Markov chain (5):

$$P(\mathbf{y'}_i, y_i | x_i) = P(\mathbf{y'}_i | x_i) P(y_i | x_i)$$

Thus:

$$P(x_i | \mathbf{y'}_i, y_i) = P(\mathbf{y'}_i | x_i) P(y_i | x_i) P(x_i) \frac{1}{P(\mathbf{y'}_i, y_i)} \qquad (22)$$

Let us denote the actually transmitted $x_i$ by $x_t$. Perfect decoding of $x_i$ implies $P(x_i | \mathbf{y'}_i, y_i) = 0$ for $x_i \neq x_t$ for all $y_i$ and $\mathbf{y'}_i$ possible when $x_i = x_t$. For the channel (1) all $y_i$ have non-zero probability for both possible $x_t$, that is $P(y_i | x_i) > 0$. Then since any of the terms on the right hand side of (22) except of the first one is not zero for all possible $\mathbf{y'}_i$ and $x_i$, the first term must be $P(\mathbf{y'}_i | x_i) = 0$, for $x_i \neq x_t$, which ensures perfect decoding of $x_i$ from $\mathbf{y'}_i$, implying (8).

## *Appendix B*

Proof of (9):

In this appendix we shall use two types channel inputs. One of them will be a codeword **x** chosen randomly and uniformly from the good code (GC) X approaching capacity within $\varepsilon'$ at channel SNR $s_0$. All the properties related to this input will be denoted by the superscript $^{GC}$, such as $I^{GC}$. The other type of input will be a vector **x** of symbols $x_i$ chosen independently and according to the symbol-vise distribution of our good code X which may be dependent to a certain extent on the symbol index i. We denote this input distribution by the superscript $^{IND}$ and the corresponding mutual information as $I^{IND}(\mathbf{x};\mathbf{y})$. The symbol-wise mutual information $I(x_i;y_i)$ is identical for both the distributions for each i. Thus the first term in (9) equals $\frac{1}{n} I^{IND}(\mathbf{x};\mathbf{y})$ and the second equals $\frac{1}{n} I^{GC}(\mathbf{x};\mathbf{y})$.

In the rest of this appendix we shall denote by $y^0$ the output of a channel with SNR equal to $s^0$ and by $y^l$ and $y^h$ the output of a channel parameterized by some

$s^l < s^h < s^0$.

For both the $\mathbf{x}^{GC}$ and $\mathbf{x}^{IND}$ types of channel inputs

$$I(\mathbf{x};\mathbf{y}) = H(\mathbf{y}) - H(\mathbf{y}|\mathbf{x}) \qquad (23)$$

Since H(**y**|**x**) is invariant with respect to the type of channel input ($^{GC}$ or $^{IND}$) over the memoryless channel, the difference DI=$I^{IND}$(**x**,**y**)- $I^{GC}$(**x**,**y**) is determined wholly by

$$DI = H^{IND}(\mathbf{y}) - H^{GC}(\mathbf{y}) \qquad (24)$$

By the chain rule of entropy we have for both the types of channel inputs:
$$H(\mathbf{y}^h,\mathbf{y}^l) = H(\mathbf{y}^l) + H(\mathbf{y}^h|\mathbf{y}^l) = H(\mathbf{y}^h) + H(\mathbf{y}^l|\mathbf{y}^h)$$

$$H(\mathbf{y}^l) = H(\mathbf{y}^h) + H(\mathbf{y}^l|\mathbf{y}^h) - H(\mathbf{y}^h|\mathbf{y}^l) \qquad (25)$$

Let us compare H($\mathbf{y}^l$) for the two types of channel inputs taking into account that H($\mathbf{y}^l$|$\mathbf{y}^h$) does not depend on the channel input type due to the Markov (2) and memoryless properties of the channel:

$$H(\mathbf{y}^l)^{IND} - H(\mathbf{y}^l)^{GC} = H(\mathbf{y}^h)^{IND} - H(\mathbf{y}^h)^{GC} - H(\mathbf{y}^h|\mathbf{y}^l)^{IND} + H(\mathbf{y}^h|\mathbf{y}^l)^{GC} \qquad (26)$$

Let us define $\beta = H(\mathbf{y}^h|\mathbf{y}^l)^{IND} - H(\mathbf{y}^h|\mathbf{y}^l)^{GC}$

Then from (24) and (26):

$$I(\mathbf{x};\mathbf{y}^l)^{IND} - I(\mathbf{x};\mathbf{y}^l)^{GC} = I(\mathbf{x};\mathbf{y}^h)^{IND} - I(\mathbf{x};\mathbf{y}^h)^{GC} - \beta \qquad (27)$$

The difference $\beta$ is positive or zero since the $^{IND}$ and the $^{GC}$ distributions induce the same symbol-wise distributions $p(y_i^l, y_i^h)$ while only the $^{GC}$ induces dependence between different symbols. Furthermore $DI = I(\mathbf{x};\mathbf{y})^{IND} - I(\mathbf{x};\mathbf{y})^{GC}$ is always positive or zero since the symbol-wise distributions of the two input types are identical while only the $^{GC}$ input induces dependency between the input symbols. Thus $DI = I(\mathbf{x};\mathbf{y})^{IND} - I(\mathbf{x};\mathbf{y})^{GC}$ is a positive non-decreasing function of s. At s= $s^0$, DI is small as desired since $\frac{1}{n}I(\mathbf{x};\mathbf{y})^{GC}$ approaches capacity within $\varepsilon'$ at channel SNR of $s^0$ while $\frac{1}{n}I(\mathbf{x};\mathbf{y})^{IND}$ cannot exceed it. This proves (9) including $\gamma(s)$ being non-decreasing function of s.

*Appendix C*

Proof of (17):

Denote the MMSE estimators (16) of $x_i$ by

$$A = P(x_i=1|y_i)=\hat{x}_i(y_i)$$
$$B = P(x_i=1|\mathbf{y'}_i,y_i)=\hat{x}_i(\mathbf{y'}_i,y_i) \quad (28)$$

Then, by (15) :

$$\varepsilon \geq E[(x-A)^2 - (x-B)^2]$$
$$= E[A^2 - B^2 - 2x(A-B)]$$
$$= E[(A-B)(A+B) - 2x(A-B)]$$
$$= E[(A-B)(A-x+B-x)]$$

$$\varepsilon \geq E[(A-B)(A-x+B-x)] \quad (29)$$

It is well known that the error of an MMSE estimator is not correlated neither to the estimate itself and neither to any function of the information which was used to form the estimate. Now, since B is the MMSE estimate using the full information **y** which can also produce A, we have 0= E[A(B-x)]= E[B(B-x)]. Then 2(B-x) can be subtracted from the term inside the right parenthesis in (29) yielding

$$\varepsilon \geq E[(A-B)(A-x-(B-x))]$$
$$\varepsilon \geq E[(A-B)(A-B)] = E(A-B)^2 \quad (30)$$

This, together with the definitions (28) implies (17).

*Appendix D*

1. *Proof of* (18)

Notation: $c_i$ denote various strictly positive and finite constants required for the various steps.

Since $\mathbf{y'}_i$ and $y_i$, are independent sources of information about $x_i$, in the sense of (5), there exist a function F, analyzed in section 2 of this appendix:

$$P(x_i =1|\mathbf{y'}_i, y_i) = F[\, P(x_i =1|y_i)\,,\, P(x_i =1|\mathbf{y'}_i)\,] \qquad (31)$$

The range of $P(x_i=1)$ is limited to a narrow range around 0.5, say 0.4 to 0.6, as explained in the paragraph below (19), that is we can disregard the cases of $P(x=1)$ deviating significantly from 0.5 since this can happen only at vanishing proportion of the coded bits, otherwise the code cannot approach capacity and vanishing proportion of coded bits cannot influence the average EI because EI is limited to one per bit in our binary transmission scheme.

Using the function F, (17) can be rewritten as

$$E\{F[P(x_i =1|y_i)\,,\, P(x_i =1|\mathbf{y'}_i)] - P(x_i =1|y_i)\}^2 \leq \varepsilon$$

Defining

$$G(y_i, \mathbf{y'}_i) \triangleq \{F[P(x_i =1|y_i)\,,\, P(x_i =1|\mathbf{y'}_i)] - P(x_i =1|y_i)\}^2$$

we have

$$E[G(y_i, \mathbf{y'}_i)] \leq \varepsilon \qquad (32)$$

Carrying the expectation over $y_i$, $x_i$ and $\mathbf{y'}_i$:

$$\sum_{x_i=0}^{1} P(x_i) \int_{\mathbf{y'}_i} d\mathbf{y'}_i [p(\mathbf{y'}_i|x_i) \int_{y_i} dy_i\, p(y_i|x_i) G(y_i, \mathbf{y'}_i)] \leq \varepsilon \qquad (33)$$

By its definition, G is non-negative. It is shown in section 2 of this appendix that if $y_i$ is limited to a certain region $\Upsilon$ which can be chosen so that

$P(\Upsilon) \triangleq \min[P(y_i \in \Upsilon | x_i = 0), P(y_i \in \Upsilon | x_i = 1)]$ is strictly positive, then G can be shown to be lower bounded by

$$G > \alpha[P(x_i=1|\mathbf{y'}_i)-P(x_i=1)]^2 \tag{34}$$

for some positive $\alpha$ independent of $\mathbf{y'}_i$. Then (33) can be lower bounded by using the last expression instead of G and by integrating $y_i$ only over $\Upsilon$. This yields:

$$\sum_{x_i=0}^{1} P(x_i) \int_{\mathbf{y'}_i} d\mathbf{y'}_i \{p(\mathbf{y'}_i | x_i)\alpha[P(x_i=1|\mathbf{y'}_i)-P(x_i=1)]^2 \int_{y_i \in \Upsilon} dy_i p(y_i | x_i)\} \leq \varepsilon$$

$$\sum_{x_i=0}^{1} P(x_i) \int_{\mathbf{y'}_i} d\mathbf{y'}_i \{p(\mathbf{y'}_i | x_i)\alpha[P(x_i=1|\mathbf{y'}_i)-P(x_i=1)]^2 P(\Upsilon)\} \leq \varepsilon$$

$$E\{\alpha[P(x_i=1|\mathbf{y'}_i)-P(x_i=1)]^2 P(\Upsilon)]\} \leq \varepsilon$$

This yields

$$E[P(x_i=1|\mathbf{y'}_i)-P(x_i=1)]^2 \leq \frac{\varepsilon}{\alpha P(\Upsilon)} \tag{35}$$

implying (18) by the following steps:

We shall drop the indices i in the following since they are not essential for the derivation.

Since $P(x=1)=1-P(x=0)$, the above equation holds for $x=0$ as well. Now

$$I(x;\mathbf{y'}) = E[\log(P(x|\mathbf{y'}) - \log(P(x))]$$

$$I(x;\mathbf{y'}) = \sum_{x=0}^{1} [\int_{\mathbf{y'}} p(\mathbf{y'})P(x|\mathbf{y'})\log(P(x|\mathbf{y'})d\mathbf{y'} - P(x)\log(P(x))] \tag{36}$$

We shall treat the $x=1$ term of the sum, the $x=0$ term is bounded by the same method.

$$I1 \triangleq \int_{\mathbf{y'}} P(\mathbf{y'})P(x=1|\mathbf{y'})\log(P(x=1|\mathbf{y'})d\mathbf{y'} - P(x=1)\log(P(x=1)$$

$$I1 = E_{\mathbf{y'}}[P(x=1|\mathbf{y'})\log(P(x=1|\mathbf{y'}) - P(x=1)\log(P(x=1))] \tag{37}$$

To streamline the presentation

denote $q = P(x=1|\mathbf{y'})$, $q_0 = P(x=1)$, $h1(q) = q\log(q)$. Then the last equation becomes

$$I1 = E_{\mathbf{y'}}[h1(q(\mathbf{y'}) - h1(q_0)] \tag{38}$$

The function (38) can be evaluated as the expectation of the vertical difference between $hl(q(\mathbf{y}'))$ and the horizontal line crossing the point $[q_0, hl(q_0)]$, see the figure below. The horizontal line may be replaced by any straight line crossing the above point without changing the result, this is equivalent to adding $c_1(q-q_0)$ to (38), which does not change the outcome because $E_{\mathbf{y}'}[P(x=1|\mathbf{y}')] = P(x=1) \rightarrow E_{\mathbf{y}'}(q) = q_0$. We shall use the line, depicted below, which is a tangent to $hl(q)$ at $q=q_0$. The function $q\log(q)$ is concave, has a positive and finite second derivative at all points of interest near $q_0=0$ and can be upper bounded by a parabola (quadratic function of q) touching it at the point $q=q_0$. This implies (38) is upper-bounded by

$I1 \leq c_2 E_{\mathbf{y}'}[P(x=1|\mathbf{y}') - P(x=1)]^2$ which, together with (35) and (36) yields (18).

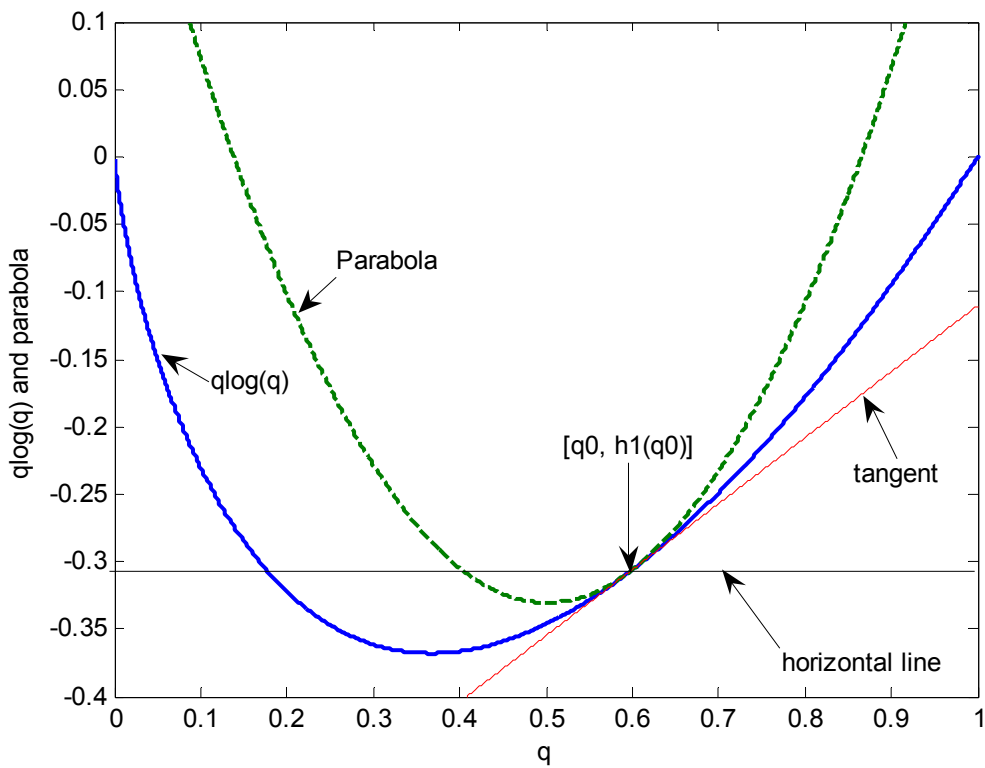

*figure 2: Plot of the qLog(q) function and it's quadratic upper-bound.*

## 2. Functions F and G

Denote

$$A = P(x_i=1|\mathbf{y'}_i)=\hat{x}_i(\mathbf{y'}_i)$$
$$B = P(x_i=1|\mathbf{y}_i)=\hat{x}_i(\mathbf{y}_i)$$
$$F = P(x_i=1|\mathbf{y'}_i,\mathbf{y}_i)=\hat{x}_i(\mathbf{y'}_i,\mathbf{y}_i) \qquad (39)$$
$$a = \mathbf{y'}_i$$
$$b = \mathbf{y}_i$$

Then:

$$F = P(x_i=1|a,b) = P(a,b|x_i=1)P(x_i=1)/P(a,b)$$
$$F = P(a|x_i=1)P(b|x_i=1)P(x_i=1)/P(a,b)$$
$$F = P(x_i=1|a)P(x_i=1|b)P(a)P(b)/[P(a,b)P(x_i=1)]$$
$$F = A \cdot B \cdot P(a)P(b)/[P(a,b)P(x_i=1)]$$

Dividing by $P(x_i=1|a,b) + P(x_i=0|a,b) = 1$ we have:

$$\frac{A \cdot B \cdot P(a)P(b)/[P(a,b)P(x_i=1)]}{A \cdot B \cdot P(a)P(b)/[P(a,b)P(x_i=1)] + (1-A)(1-B)P(a)P(b)/[P(a,b)(1-P(x_i=1))]}$$

$$F = \frac{A \cdot B(1-P(x_i=1))}{AB(1-P(x_i=1)) + (1-A)(1-B)P(x_i=1)} \qquad (40)$$

Set $K = \dfrac{1-P(x_i=1)}{P(x_i=1)}$. This cannot deviate too much from 1 as stated above.

Let us examine G

$$G = (F-B)^2 =$$
$$G = [\frac{ABK}{ABK + (1-A)(1-B)} - B]^2 =$$
$$G = [\frac{B^2 + (1+K)AB - B - (1+K)AB^2}{(1+K)AB + 1 - A - B}]^2$$
$$G = [\frac{[(1+K)A-1]B - [(1+K)A-1]B^2}{(1+K)AB + 1 - A - B}]^2$$

$$G = [\frac{[(1+K)A-1]B(1-B)}{(1+K)AB+1-A-B}]^2$$

$$G = [\frac{(1+K)(A-\frac{1}{(1+K)})B(1-B)}{(1+K)AB+1-A-B}]^2$$

Using $P(x=1) = \frac{1}{(1+K)}$ we have:

$$G = [\frac{(1+K)(A-P(x=1))B(1-B)}{(1+K)AB+1-A-B}]^2$$

$$G = (A-P(x=1))^2 [\frac{(1+K)}{(1+K)AB+1-A-B}]^2 [B(1-B)]^2 \quad (41)$$

The values of A and B are between 0 to 1 and K is nearly 1 as explained above. Furthermore we shall limit the range of $y_i$ as to limit B to a range say, 0.1 to 0.9. Then the second term of (41) is

$$\frac{(1+K)}{(1+K)AB+1-A-B} =$$
$$\frac{(1+K)}{KAB+AB+1-A-B}$$
$$= \frac{1+K}{KAB+(1-A)(1-B)}$$

which is a strictly positive number of limited range and so is the $B(1-B)$ term in (41). This, together with (41) results in $G \geq c_3(A-P(x=1))^2$. This together with the definition of A in (39) implies (34).

*References:*

[1] A. Sanderovich, M. Peleg, and S. Shamai (Shitz): " LDPC Coded MIMO Multiple Access with Iterative Joint Decoding ", *IEEE Transactions on Information Theory*, Volume: 51, Issue: 4, pp. 1437-1450, April 2005

[2] S. ten Brink: "Designing Iterative Decoding Schemes with the Extrinsic Information Transfer Chart", *AEU International Journal of Electronics and Communications*, vol. 54 no. 6, Dec. 2000, pp. 389-398.

[3] M. Peleg and S. Shamai (Shitz): "Efficient Communication over the Discrete-Time Memoryless Rayleigh Fading Channel with Turbo Coding/Decoding", *European Transaction on Telecommunications (ETT)*, vol. 11, Sept/Oct 2000, pp. 475-485.

[4] C. Berrou, A. Glavieux and P. Thitimajshima, "Near Shannon limit error-correcting coding and decoding," *Proceedings of Int. Communications Conf. (ICC'93)* Geneva, Switzerland, May 1993, pp. 1064-1070.

[5] T.S. Han and S. Verdu, "Approximation theory of output statistics," *IEEE Transactions on Information Theory*, Volume: 39, Issue: 3, May 1993, pp. 752 – 772.

[6] D. Guo, S. Shamai (Shitz) and S. Verdu, ``Mutual Information and Minimum Mean-Square Error in Gaussian Channels'' , *IEEE Transactions on Information Theory*, Volume: 51, Issue: 4, pp. 1261-1282 April 2005

[7] I. Sutskover, S. Shamai (Shitz) and J. Ziv,``Extremes of Information Combining'' , *IEEE Transactions on Information Theory*, Volume: 51, Issue: 4, pp. 1313-1325, April 2005

[8] S. Shamai (Shitz) and S. Verdu, "The Empirical Distribution of Good Codes", *IEEE Trans. on Information Theory*, Vol. 43, No. 3, pp. 836-846, May 1997.

[9] A. Ashikhmin, G. Kramer, S. ten Brink: "Code Rate and the Area under Extrinsic Information Transfer Curves", Proceedings of the *International Symposium on Information Theory*, 2002, p.115, IEEE

[10] Part 2 of this report : "On Extrinsic Information of Good Codes Operating Over Discrete Memoryless Channels with Incremental Noisiness."

[11] C. Meassone, R. Urbanke, A. Montanari, and T. Richardson: "Life Above Threshold: From List Decoding to Area Theorem and MSE", *IEEE Information Theory Workshop ( ITW)*, Oct. 24-29 2004, Texas.

# Part 2

# On Extrinsic Information of Good Codes Operating Over Discrete Memoryless Channels with Incremental Noisiness.

M. Peleg, A. Sanderovich and S. Shamai (Shitz)

*Abstract:* We show that the Extrinsic Information about the coded bits of any good (capacity achieving) code operating over a wide class of discrete memoryless channels (DMC) is zero when channel capacity is below the code rate and positive constant otherwise, that is, the Extrinsic Information Transfer (EXIT) chart is a step function of channel quality, for any capacity achieving code. The results are proved for the binary symmetric channel and for the binary erasure channel while proof for additional channels with incremental noisiness such as the AWGN channels with QAM and inputs requires further elaboration. It follows that, for a common class of iterative receivers where the error correcting decoder must operate at first iteration at rate above capacity (such as in turbo equalization, turbo channel estimation, parallel and serial concatenated coding and the like), classical good codes which achieve capacity over the DMC are not effective and should be replaced by different new ones.

*I Introduction*

We derive the Extrinsic Information Transfer (EXIT) chart of asymptotically long codes which achieve a vanishing probability of error over a wide class of discrete memoryless channels (DMC) when channel capacity is higher then the code rate. We denote such codes as "good codes" in the following. In our derivation we use discrete channel model, however the size of the output alphabet is not limited so the results apply for example to the QPSK Gaussian Additive Channel with a fine quantizer at the receiver input. The results provide an insight about corresponding iterative receivers designed to approach the channel capacity assuming asymptotically long codewords.

It is well known that when a good Error Correcting Code (ECC) is used to transmit information over a channel the capacity of which is lower than the code rate then the error rate is high. This scenario actually occurs at the first decoding iteration performed by the new iterative receivers based on the turbo principle, see fig.1, where some preprocessor such as equalizer, multi-user detector [1], phase estimator or other precedes the decoder and employs the decoder outputs to improve the channel presented to the decoder over the successive iterations. The ECC code may be any code, including also an iteratively decodable turbo or LDPC code.

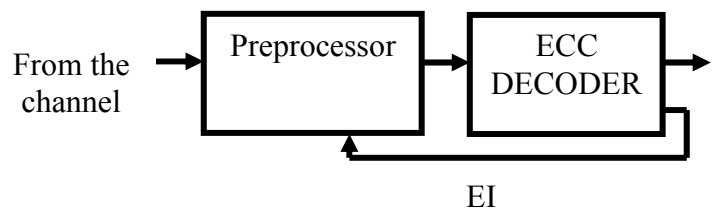

Figure 1. Iterative receiver

If the whole iterative receiver is designed to approach capacity and if the iterative feedback to the preprocessor is really required then at the first iteration the ECC decoder is presented with a channel the capacity of which is below the code rate while at the next iterations the preprocessor will improve the ECC decoder input and enable errorless decoding. To achieve this, the ECC decoder must pass some useful information to the preprocessor at the first iteration, while operating over a channel the capacity of which is below the code rate. The relevant information to be passed is the well known Extrinsic Information (EI) to be defined below.

In fact, serially concatenated turbo codes can also be represented by the structure of fig.1 where the preprocessor is the decoder of the inner component code. Parallel concatenated turbo codes are decoded by a similar structure and the operation at rate above channel capacity at first iteration is then also clearly required since the component code is presented only with a subset of the channel output symbols.

In the following derivation we will show that over a wide class of DMC, codes considered to be good in the classical sense, provide no EI at all in this setting and thus eliminate any improvement by the iterative feedback in fig. 1. So the search for different codes fitting the new iterative systems such as performed in [1], [2],[3] and others is indeed essential if the iterative receiver is to perform better than, say, separate equalization and decoding. Also, classical good codes cannot perform well as outer codes in a serially concatenated turbo code or component codes in parallel concatenated turbo codes. This was indicated first in [4] where increasing the constraint length of a convolutional component code rendered the iterative feedback of a turbo decoder ineffective. More precisely, we will show that the Extrinsic Information Transfer (EXIT) chart of any good code over the M-ary input channel is a step function, the EI being zero for channel quality at which the channel capacity is below the code rate and log(M) at channel quality at which the capacity exceeds the code rate where M denotes the size of the alphabet at the channel input and all logarithms here are taken with base 2. The results require the channel quality to be determined by a single parameter w such as SNR in a manner we denoted 'incremental noisiness'. This ensures that the code remain matched well enough to the channel over the range of channel qualities. Notations: Mutual information is denoted by I, entropy by H, P and p denote probability and a probability density function respectively and bold letters denote vectors.

## *II Models and definitions*

1. *Channel model*:

We examine a time invariant DMC with input alphabet of size M characterized by $p(y|x)$ where and x and y are the input and the output symbols of the channel respectively. C denotes the channel capacity.

We shall restrict our treatment to channels having the following properties denoted collectively as incremental noisiness:

The channel transfer function P(y|x) is determined by a single parameter w denoted as noisiness, such as 1/SNR for the AWGN channel or Bit Error Rate over the Binary Symmetric Channel (BSC).

A channel characterized by $w^h$, for some $w^h > w^l$, can be described as physically degraded with respect to a channel characterized by $w^l$, where the superscripts h and l denote 'higher' and 'lower' noisiness respectively. The outputs of the two channels are denoted $y^h$ and $y^l$. By physically degraded we mean (as in [9]) that $x_i, y_i^l, y_i^h$ form the Markov chain:

$$x_i - y_i^l - y_i^h \tag{1}$$

We shall let the outputs y of our incremental noisiness channel to be governed by the transition rate matrix $A = \{a_{kl}\}$ in the same manner as the discrete states of the discrete, continuous time Markov process but with time replaced by $w_i$, so that

$$\frac{d}{dw_i} P(y_k) = \sum_l a_{kl} P(y_l) \quad \text{where } P(y_l) \triangleq P(y = y_l)$$

$$a_{ll} = -\sum_{k \neq l} a_{kl} \tag{2}$$

$$k \neq l \rightarrow a_{kl} > 0$$

The requirement of all strictly positive $a_{kl}$ can be partly released as explained below (26), that is the positive $a_{kl}$ need only to provide a path with nodes $y_i$ and edges $a_{kl}$ of the form $y_1$-$a_{12}$-$y_2$-$a_{26}$-$y_6$-… from any $y_l$, $p(y_l) > 0$ to any $y_k$, $p(y_l) > 0$ using $y_i$, $p(y_i) > 0$ as intermediate nodes.

Many channels conform to those restrictions, for example the the binary symmetric channel (BSC) and Binary Erasure Channel (BEC). Also the finely quantized AWGN channel and Raleigh and Rice fading channels with additive Gaussian noise with discrete input alphabet can be modeled well by this framework. The model of incremental noisiness is essential not only to our arguments but also to the definition of a good code used in this paper.

In order to perform particial derivatives in our derivation, we shall need an extension to a channel characterized for each channel use i by a different $w_i$, so we use $\mathbf{w}=[w_1,w_2\ldots w_i..w_N]$. Non-bold w will mean that all $w_i$ are equal to w. Derivative with respect to $w_i$ with the other elements of $\mathbf{w}$ held constant is denoted as $\frac{d()}{dw_i}$, while $\frac{d()}{dw}$ denotes a derivative with respect to the channel quality w common to all the elements of $\mathbf{w}$.

An additional requirement from the channel is 'sufficient transparency' (ST), that is maintaining some minimal influence of the input over the outputs of the channel as defined below as follows: Define the vector $\mathbf{Vy}$ as comprising the M elements $\mathbf{Vy}(m)=P(y\,|\,x=x_m)$. It is required that some set of m values of y can be found which produces m independent vectors $\mathbf{Vy}$.

*2. The good code:*

We desire to transmit information $\mathbf{U}$. We do so in the standard manner of transmitting a codeword $\mathbf{x}$, a vector of n channel symbols x belonging to an asymptotically long good code X of rate R. The selection of the transmitted codewords is determined by $\mathbf{U}$ and is equi-probable. The received vector is denoted $\mathbf{y}$. The code is such that it achieves vanishing error probability at channel noisiness level $w_0$ for which $R>C-\varepsilon'$ for any small positive $\varepsilon'$ when n is large enough. The Markov chain (1) implies vanishing error probability also for all $w<w_0$.

*3. The Extrinsic Information:*

We are interested in a symbol $x_i$, which is a symbol at the i'th position in **x**. We define **x**$_i$' as **x** with $x_i$ excluded and correspondingly **y**$_i$' as **y** with $y_i$ excluded.

We denote $z_i$ the complete information obtainable from **y**$_i$' about $x_i$, known as the extrinsic information, see for example [4] and [5]. The extrinsic information $z_i$ can be expressed for example over the binary channel as the Logarithmic Likelihood Ratio (LLR) of $x_i$ and over the M-ary channel as an M-tuple of probabilities of $x_i$ being equal to the possible M channel inputs.

The average extrinsic information measure is then defined as

$$\text{EI} = \frac{1}{n}\sum_{i=1}^{n} I(x_i; \mathbf{y}'_i) = \frac{1}{n}\sum_{i=1}^{n} I(x_i; z_i) \qquad (3)$$

When $x_i$ is given, then $y_i$ is independent of **y**$_i$', that is $P(y_i | x_i, \mathbf{y}'_i) = P(y_i | x_i)$ as shown already in [4]. This extends the Markov chain (1) to

$$\mathbf{y}'^h_i - \mathbf{y}'^l_i - x_i - y^l_i - y^h_i \qquad (4)$$
$$z_i - x_i - y^l_i - y^h_i \qquad (5)$$

Furthermore, due to this Markov chain and the data processing theorem we have $I(\mathbf{y}'^l_i; x_i) \geq I(\mathbf{y}'^h_i; x_i)$, thus EI is a non-increasing function of w.

$$EI(w^l) \geq EI(w^h) \qquad (6)$$

*III EXIT chart of good codes*

When the code rate R is strictly below the capacity C we have perfect decoding for asymptotically long good codes, even if the single symbol $y_i$ is removed (erased) before the decoding, thus we have

$$R < C \Rightarrow EI = \log(M) \qquad (7)$$

This intuitive attribute of good codes is proved in Appendix A under the mild condition that the channel is noisy so that $x_i$ cannot be decoded with zero error probability from $y_i$ alone, that is $p(x_i|y_i)>0$ for some values of $x_i$ different from the one which was transmitted.

The central result of this work is the following proposition and its method of proof:

*Proposition: The average EI which can be obtained from all the channel outputs about the coded bits $x_i$ of a good code operating over a discrete memoryless channel with incremental noisiness and sufficient transparency, the capacity of which is below the code rate, is zero.*

The proof is exact for the BSC and BEC channels and needs further elaboration for finely quantized channels such as the AWGN channel.

Proof:

We shall examine a code which is capacity achieving for some noisiness $w=w_0$, that is, it is capable of reliably transmitting information at a rate $R = C(w_0) - \varepsilon'$ over a channel parameterized by $w=w_0$ for any small $\varepsilon'$.

Our proof uses the concept of GEXIT as introduced in [5]. GEXIT is defined [5, eq.(2)] as:

$$GEXIT = \frac{1}{n}\frac{d}{dw}H(\mathbf{x}|\mathbf{y}) \qquad (8)$$

which is transformed easily to:

$$GEXIT = \frac{1}{n}\frac{d}{dw}[H(\mathbf{x}) - I(\mathbf{x};\mathbf{y})]$$
$$GEXIT = -\frac{1}{n}\frac{d}{dw}I(\mathbf{x};\mathbf{y}) \qquad (9)$$

GEXIT is an average of $GEXIT_i$, [5, eq.(3)]:

$$GEXIT = \frac{1}{n}\sum GEXIT_i$$
$$GEXIT_i = \frac{d}{dw_i}H(\mathbf{x}|\mathbf{y}) \qquad (10)$$

It was shown in [5, eq.(5)] that:
$$GEXIT_i = \frac{d}{dw_i} H(x_i | z_i, y_i) \qquad (11)$$

Similarly to (9), GEXIT$_i$ from (11) is
$$GEXIT_i = -\frac{d}{dw_i} I(x_i; z_i, y_i) \qquad (12)$$

Now let us introduce GEXIT0 and GEXIT0$_i$ defined as GEXIT and GEXIT$_i$ respectively but with transmitted symbols x0$_i$ distributed independently, which implies z$_i$=0, and still according to the symbol-vise distribution of our good code X which may be dependent to a certain limited extent [8] on the symbol index i. We denote this input distribution by the superscript $^{IND}$ and the corresponding mutual information as $I^{IND}$(**x**0;**y**0). Of course (12) applies, so $GEXIT0_i = -\frac{d}{dw_i} I(x0_i; y0_i)$.

Now let us compare GEXIT$_i$ to GEXIT0$_i$. We shall use (12) and the fact that x$_i$ are distributed identically to x0$_i$ :

$$GEXIT0_i - GEXIT_i = -\frac{d}{dw_i} I(x_i; y_i) + \frac{d}{dw_i} I(x_i; z_i, y_i)$$
$$= \frac{d}{dw_i}\left[I(x_i; z_i, y_i) - I(x_i; y_i)\right]$$

By the chain rule of mutual information we have then:

$$GEXIT0_i - GEXIT_i = \frac{d}{dw_i}\left[I(x_i; z_i | y_i)\right]$$
$$= \frac{d}{dw_i}\left[I(z_i; x_i, y_i) - I(z_i; y_i)\right]$$

By the Markov chain (5) the first I( ) term is not dependent on y$_i$ and we have:
$$GEXIT0_i - GEXIT_i = -\frac{d}{dw_i} I(z_i; y_i) \qquad (13)$$

It is well known that, when the code rate R is at or above capacity, good codes mimic closely the channel output statistics of a capacity achieving identically and independently distributed ( i.i.d.) input [6, Theorem 15].

Specifically, we prove in Appendix B that for w>$w_0$ the mutual information I (**x;y**) over the channel with the good code is similar to $I^{IND}$(**x0;y0**). That is for any small $\varepsilon \geq 0$ and sufficiently large n:

$$0 \leq \frac{1}{n} I^{IND}[\mathbf{x}_0; \mathbf{y}_0(w)] - \frac{1}{n} I[\mathbf{x}; \mathbf{y}(w)] = \gamma(w) \leq \varepsilon^3 = \varepsilon' \qquad (14)$$

and $\gamma$ is non-increasing function of w. The substitution $\varepsilon' = \varepsilon^3$ will be required below.

The end of appendix B conjectures that if the channel is symmetric then both the above mutual information terms approach capacity, however pursuing this is not needed here and is out of scope of this work.

It follows from (14) that for any pair of values w1 < w2 for which R>C we have:

$$0 \leq I^{IND}[\mathbf{x}_0; \mathbf{y}_0(w_1)] - I^{IND}[\mathbf{x}_0; \mathbf{y}_0(w_2)] - \{I[\mathbf{x}; \mathbf{y}(w_1)] - I[\mathbf{x}; \mathbf{y}(w_2)]\} = n\gamma(w_1) - n\gamma(w_2) \leq n\varepsilon^3$$

Applying (9) we get

$$0 \leq \int_{w_1}^{w_2} [GEXIT0(w) - GEXIT(w)] dw \leq \gamma(w_1) - \gamma(w_2) \leq \varepsilon^3 \qquad (15)$$

also, from (13) and the Markov chain (5):

$$DG(w) \triangleq GEXIT0(w) - GEXIT(w) \geq 0 \qquad (16)$$

Now lets place $w_1$ at the threshold value $w_1=w_0$. Since $\gamma(w)$ is non-increasing and DG is non-negative then its average over an interval of $w_1=w_0$ $to$ $w_2 = w_0 + \Delta$ cannot exceed $\varepsilon^3/\Delta$, otherwise its integral (15) would exceed $\varepsilon^3$. We shall choose $\Delta = \varepsilon$ to limit the average DG to $\varepsilon^2$. Thus there is some w=$w_t$ in the above interval, $\varepsilon$ within $w_0$, for which the absolute value of ID is bounded by

$$DG(w_t) \leq \varepsilon^2 \qquad (17)$$

Due to (6), vanishing EI at $w_t$ implies vanishing EI at all larger values of w, so it is sufficient to prove vanishing EI at $w_t$.

It follows from (13), (16) and (17) that:

$$0 \geq \frac{1}{n}\sum_i \frac{d}{dw_i} I(z_i; y_i) \geq -\varepsilon^2 \qquad (18)$$

By the Markov chain (5) each element of the above sum is negative and their average is lower bounded by $-\varepsilon^2$. This implies that each element is lower bounded by $-\varepsilon$, except at most $n\varepsilon$ elements which may be more negative (with $n\varepsilon$ elements more negative than $-\varepsilon$, (18) will be violated). This vanishing proportion of elements can contribute only $\varepsilon \log(M)$ bits to the average EI, eq. (3), because the EI for each bit is bounded by log(M), so we can disregard them in our proof of vanishing average EI and use

$$0 \geq \frac{d}{dw_i} I(z_i; y_i) \geq -\varepsilon \qquad (19)$$

It is shown in [7] that GEXIT determines exactly the Minimum Mean Squared Error over Gaussian channels. This provides means to prove our results over the AWGN channel, see [11], however here we shall pursue a different approach to provide a more general result.

Now:

$$\frac{d}{dw_i} I(z_i; y_i) = \frac{-d}{dw_i} H(z_i | y_i)$$

$$\frac{d}{dw_i} I(z_i; y_i) = \frac{d}{dw_i} \int dz_i \sum_{y_i} p(z_i, y_i) \log p(z_i | y_i)$$

$$\frac{d}{dw_i} I(z_i; y_i) = \int dz_i \sum_{y_i} \log p(z_i | y_i) \frac{d}{dw_i} p(z_i, y_i) + \int dz_i \sum_{y_i} \frac{p(z_i, y_i)}{p(z_i | y_i)} \frac{d}{dw_i} p(z_i | y_i)$$

$$\frac{d}{dw_i} I(z_i; y_i) = \int dz_i \sum_{y_i} \log p(z_i | y_i) \frac{d}{dw_i} p(z_i, y_i) + \sum_{y_i} p(y_i) \frac{d}{dw_i} \int dz_i p(z_i | y_i)$$

$$\frac{d}{dw_i} I(z_i; y_i) = \int dz_i \sum_{y_i} \log p(z_i | y_i) \frac{d}{dw_i} p(z_i, y_i) \qquad (20)$$

By (2) and by the Markov chain (5) we have:

$$\frac{d}{dw_i} p(y_k, z) = \sum_l a_{kl} P(y_l, z) \qquad (21)$$

Using this in (20) while dropping most of the indices i for clarity we get:

$$\frac{d}{dw_i} I(z; y) = \int dz \sum_y \log p(z | y) \frac{d}{dw_i} p(z, y_i)$$

$$\frac{d}{dw_i} I(z; y) = \int dz \sum_k \log p(z | y_k) \sum_l a_{kl} p(z, y_l)$$

$$\frac{d}{dw_i} I(z; y) = \sum_l \sum_k \int dz \log p(z | y_k) a_{kl} p(z, y_l)$$

$$\frac{d}{dw_i} I(z; y) = \sum_l P(y_l) \sum_k \int dz \, a_{kl} p(z | y_l) \log p(z | y_k)$$

$$\frac{d}{dw_i} I(z; y) = \sum_l P(y_l) \sum_{k \neq l} \int dz \, a_{kl} p(z | y_l) [\log p(z | y_k) - \log p(z | y_l)]$$

$$\frac{d}{dw_i} I(z; y) = -\sum_l P(y_l) \sum_{k \neq l} a_{kl} D[p(z | y_l) \| (z | y_k)] \qquad (22)$$

The line before last came from $a_{ll} = -\sum_{k \neq l} a_{kl}$ in (2) and D[.||.] denotes the Kullback-Leibler divergence. Since D[.||.] is non-negative, (22) and (19) imply

$$0 \leq D[p(z|y_l) \| (z|y_k)] \leq \frac{\varepsilon}{P(y_l)a_{kl}} \quad (23)$$

The last expression implies

$$p(z|y=y_l) = (z|y=y_k) \quad (24)$$

for all combinations off $p(y=y_l) > 0$ and $a_{kl} > 0$

This holds accurately for the BSC and BEC for which the denominator of (23) is well defined and needs further elaboration for finely quantized channels which is not performed in this work.

Our channel definition includes the condition that positive $a_{kl}$ provide a path with nodes $y_k$ and edges $a_{kl}$ of the form $y_1$-$a_{12}$-$y_2$-$a_{26}$-$y_6$-… from any $y_l$, $p(y_l)>0$ to any $y_k$, $p(y_l)>0$ using $y_k$, $p(y_k)>0$ as intermediate nodes. So (24) holds for any pair k,l, thus

$$P(z|y) = P(z)$$

So y is independent of z and:
$$P(y|z) = P(y) \quad (25)$$

Now

$$P(y|z) = \sum_{m=1}^{M} P(y|z, x=x_m)P(x=x_m|z) = \sum_{m=1}^{M} P(y|x=x_m)P(x=x_m|z)$$

and

$$P(y) = \sum_{m=1}^{M} P(y|x=x_m)P(x=x_m)$$

By substituting the last two equations into (25) we have:

$$\sum_{m=1}^{M} P(y|x=x_m) \cdot [P(x=x_m|z) - P(x=x_m)] = 0 \quad (26)$$

The sum can be regarded as an inner product of two m-dimensional vectors **Vy** and **Vz** with elements **Vy**$(m) = P(y|x=x_m)$ and **Vz**$(m) = P(x=x_m|z) - P(x=x_m)$ respectively. The independence of z and x, $P(x=x_m|z) = P(x=x_m)$ will force **Vz**=0, conforming to (26). If a set of m values of y can be found which produces a set of m independent vectors **Vy**, then $P(z|x=x_m)=P(z)$ is the only possible solution to (26) yielding EI=0.

Thus the existence of a set of m values of y with p(y)>0 which produces m independent vectors **Vy** is sufficient to ensure EI=0. We denote this attribute of the channel as 'sufficient transparency' (ST). We conjecture that most channels with output alphabet larger than the input alphabet conform to the ST property. Indeed the binary input channels (binary symmetric channel, binary erasure channel, binary input AWGN, Rice and Raileigh channels) do so by a simple inspection. In Appendix C we demonstrate this property also for the corresponding channels with QPSK inputs.

Remarks: At decreasing channel noisiness which brings the channel capacity to a value above the code rate there is a sharp transition since I(**x**,**y**) is not determined by (14) any more but reaches a plateau at the code rate and the GEXIT, which is proportional to the derivative of I(**x**,**y**) with respect to w by(9), goes to zero abruptly. This transition takes place over a small region of w for which the term (14) is very small but, significantly, not zero. A similar transition of $\frac{1}{n}\sum_{i=1}^{n} I(y_i; \mathbf{y}')$, which is upper bounded by the average EI because of the Markov chain (5), occurs in the same region.

Clearly the area A under the EI versus $I(x_i;y_i)$ curve, $A = \int_0^1 EI \, d[I(x_i; y_i)]$, equals (1-R)logM, thus the step function EXIT chart derived here for good code over the M-ary discrete memoryless channel conforms to the "EXIT chart area property" of outer codes which was proved in [10] for any code over the binary erasure channel.

## *IV Conclusions*:

The EXIT chart of any good (capacity achieving) code operating over a wide class of M-ary input DMC is a step function of the channel noisiness, zero when channel capacity is below the code rate and logM at C>R . The results are proved for the BEC, and BSC while other finely quantized channels with incremental noisiness such as the QAM input AWGN channel need further elaboration. Thus codes good over memoryless channels are very inefficient when used in an iterative receiver of the type presented in figure 1 which includes turbo-equalization, iterative multi-user receivers and serially concatenated codes as special cases. Interestingly, the step function EXIT chart derived here conforms to the EXIT chart area property derived in [10] for the erasure channel.

Furthermore vanishing EI, as defined in (3), implies that good code operating at rate above channel capacity falls apart into its individual transmitted symbols in the sense that all the information about a coded bit $x_i$ is contained in the corresponding received symbol $y_i$ and no information about $x_i$ can be inferred from the other received symbols $\mathbf{y'}_i$.

*Appendix A, Extrinsic information at rates below capacity*

We will show in the following that when R<C, then (7) holds. When R<C, the symbol $x_i$ is decoded with zero error probability. By the Markov chain (1):

$$P(\mathbf{y'}_i, y_i | x_i) = P(\mathbf{y'}_i | x_i) P(y_i | x_i) \qquad (27)$$

Thus:

$$P(x_i | \mathbf{y'}_i, y_i) = P(\mathbf{y'}_i | x_i) P(y_i | x_i) P(x_i) \frac{1}{P(\mathbf{y'}_i, y_i)} \qquad (28)$$

Let us denote the actually transmitted $x_i$ by $x_m$. Perfect decoding of $x_i$ implies $P(x_i | \mathbf{y'}_i, y_i) = 0$ for any $x_i \neq x_m$ for all $y_i$ and $\mathbf{y'}_i$ possible when $x_i = x_m$. For some $x_i \neq x_m$ there must be a possible $P(y_i | x_i) > 0$ otherwise $x_i$ could be decoded perfectly

from $y_i$ alone. Then since any of the terms on the right hand side of (28) except of the first one is not zero for all possible **y'**$_i$ and $x_i$ we must have

$P(\mathbf{y'}_i | x_i) = 0$, for any $x_i \neq x_m$ when $x_m$ was transmitted, which ensures perfect decoding of $x_i$ from **y'**$_i$, implying (7).

## Appendix B

Proof of (14):

In this appendix we shall examine two types channel inputs. One of them will be a codeword **x** chosen randomly and uniformly from the good code X approaching capacity within $\varepsilon'$ at channel noisiness $w_0$. All the properties related to this input will be denoted by the superscript $^{GC}$, such as $I^{GC}$. The other type of input will be a vector **x** of symbols x chosen independently from the symbolwise distribution denoted by the superscript $^{IND}$ corresponding to GEXIT0 and defined below (12). The symbol-wise mutual information $I(x_i;y_i)$ is identical for both the distributions for each i. Thus the first term in (14) equals $\frac{1}{n}I^{IND}(\mathbf{x};\mathbf{y})$ and the second equals $\frac{1}{n}I^{GC}(\mathbf{x};\mathbf{y})$.

In the rest of this appendix we shall denote by $y^0$ the output of a channel with noisiness $w_0$ and by $y^l$ and $y^h$ the output of a channel parameterized by some $w^h > w^l > w_0$. By our channel definition the channel input and the outputs form the Markov chain $x$-$y^0$-$y^l$-$y^h$.

For both the $\mathbf{x}^{GC}$ and $\mathbf{x}^{IND}$ types of channel inputs

$$I(\mathbf{x};\mathbf{y}) = H(\mathbf{y}) - H(\mathbf{y}|\mathbf{x}) \tag{29}$$

Since $H(\mathbf{y}|\mathbf{x})$ is invariant with respect to the type of channel input ($^{GC}$ or $^{IND}$) over the DMC, the difference DI=$I^{IND}(\mathbf{x},\mathbf{y})$- $I^{GC}(\mathbf{x},\mathbf{y})$ is determined wholly by $H(\mathbf{y})$.

By the chain rule of entropy we have for both the types of channel inputs:

$$H(\mathbf{y}^h, \mathbf{y}^l) = H(\mathbf{y}^l) + H(\mathbf{y}^h \mid \mathbf{y}^l) = H(\mathbf{y}^h) + H(\mathbf{y}^l \mid \mathbf{y}^h)$$
$$H(\mathbf{y}^h) = H(\mathbf{y}^l) + H(\mathbf{y}^h \mid \mathbf{y}^l) - H(\mathbf{y}^l \mid \mathbf{y}^h)$$

Let us compare $H(\mathbf{y}^h)$ for the two types of channel inputs taking into account that $H(\mathbf{y}^h|\mathbf{y}^l)$ does not depend on the channel input type due to the Markov and memoryless properties of the channel:

$$H(\mathbf{y}^h)^{IND} - H(\mathbf{y}^h)^{GC} = H(\mathbf{y}^l)^{IND} - H(\mathbf{y}^l)^{GC} - H(\mathbf{y}^l \mid \mathbf{y}^h)^{IND} + H(\mathbf{y}^l \mid \mathbf{y}^h)^{GC} \quad (30)$$

Let us define $\beta = H(\mathbf{y}^l \mid \mathbf{y}^h)^{IND} - H(\mathbf{y}^l \mid \mathbf{y}^h)^{GC}$

Then from (29) and the sentence below it:

$$I(\mathbf{x}; \mathbf{y}^h)^{IND} - I(\mathbf{x}; \mathbf{y}^h)^{GC} = I(\mathbf{x}; \mathbf{y}^l)^{IND} - I(\mathbf{x}; \mathbf{y}^l)^{GC} - \beta \quad (31)$$

The difference $\beta$ is positive or zero since the $^{IND}$ and the $^{GC}$ distributions induce the same symbolwise distributions $P(y_i^l, y_i^h)$ while only the $^{GC}$ induces dependence between different symbols. Furthermore $DI = I(\mathbf{x}; \mathbf{y})^{IND} - I(\mathbf{x}; \mathbf{y})^{GC}$ is always positive since the symbolwise distributions of the two input types are identical while only the $^{GC}$ input may induce dependency between the input symbols and the channel is memoryless. Thus $DI = I(\mathbf{x}; \mathbf{y})^{IND} - I(\mathbf{x}; \mathbf{y})^{GC}$ is a positive non-increasing function of w. At w=w0, DI is small as desired since $\frac{1}{n} I(\mathbf{x}; \mathbf{y})^{GC}$ approaches capacity within $\varepsilon'$ at channel noisiness $w_0$ while $\frac{1}{n} I(\mathbf{x}; \mathbf{y})^{IND}$ cannot exceed it. This proves (14) including the monotonic behavior of $\gamma$.

Remark: Over symmetric channels a similar relationship can be demonstrated for $DI_1 = I(\mathbf{x}; \mathbf{y})^{CA} - I(\mathbf{x}; \mathbf{y})^{IND}$ where the superscript $^{CA}$ denotes the capacity achieving distribution and independent symbols $x_i$ following very similar arguments, particularly $H(\mathbf{y}|\mathbf{x})$ below (29) is invariant with respect to the type of channel input due to the channel symmetry, $\beta$ is positive because the capacity achieving distribution is uniform and $DI_1$ clearly must be positive or zero. This implies that for symmetric channels $I(\mathbf{x},\mathbf{y})$ approaches capacity for w>w0, however this result is not required in this work.

## Appendix C, ST property of the QPSK input channels

This appendix proves the sufficient transparency (ST) property for the QPSK input channels.

Let the possible channel inputs $x_m$, m=1,2,3,4 be {1, j, -1, -j}. Let us choose the values of y, $y_i$, i=1,2,3,4 be {0, 1, j, $(-1+j)/\sqrt{2}$ }. Then the (row) vectors **Vy** are of the form

$$\mathbf{Vy}_1 = [k \quad k \quad k \quad k]$$
$$\mathbf{Vy}_2 = [a \quad b \quad c \quad b]$$
$$\mathbf{Vy}_3 = [b \quad a \quad b \quad c]$$
$$\mathbf{Vy}_4 = [y \quad x \quad x \quad y]$$

and
$$a > b > c > 0$$
$$x > y > 0$$

This will hold for AWGN channel as well as for the Raleigh and Rice fading channels. It is easy to see that the four vectors are independent.

*References:*

[1] A. Sanderovich, M. Peleg, and S. Shamai (Shitz): " LDPC Coded MIMO Multiple Access with Iterative Joint Decoding ", *IEEE Transactions on Information Theory*, Volume: 51, Issue: 4, pp.1437-1450, April 2005

[2] S. ten Brink: "Designing Iterative Decoding Schemes with the Extrinsic Information Transfer Chart", *AEU International Journal of Electronics and Communications*, vol. 54 no. 6, Dec. 2000, pp. 389-398.

[3] M. Peleg and S. Shamai (Shitz): "Efficient Communication over the Discrete-Time Memoryless Rayleigh Fading Channel with Turbo Coding/Decoding", *European Transaction on Telecommunications (ETT)*, vol. 11, Sept/Oct 2000, pp. 475-485.

[4] C. Berrou, A. Glavieux and P. Thitimajshima, "Near Shannon limit error-correcting coding and decoding," *Proceedings of Int. Communications Conf. (ICC'93)* Geneva, Switzerland, May 1993, pp. 1064-1070.

[5] C. Meassone, R. Urbanke, A. Montanari, and T. Richardson: "Life Above Threshold: From List Decoding to Area Theorem and MSE", *IEEE Information Theory Workshop ( ITW)*, Oct. 24-29 2004, Texas.

[6] T.S. Han and S. Verdu, "Approximation theory of output statistics," *IEEE Transactions on Information Theory*, Volume: 39, Issue: 3, May 1993, pp. 752 – 772.

[7] D. Guo, S. Shamai (Shitz) and S. Verdu, ``Mutual Information and Minimum Mean-Square Error in Gaussian Channels'', *IEEE Transactions on Information Theory*, Volume: 51, Issue: 4, pp. 1261-1282 April 2005

[8] S. Shamai (Shitz) and S. Verdu, "The Empirical Distribution of Good Codes", *IEEE Trans. on Information Theory*, Vol. 43, No. 3, pp. 836-846, May 1997.

[9] T.M. Cover: "Comments on Broadcast Channels" *IEEE Trans. on Information Theory*, Vol. 44, No.6, , Oct 1998, pp. 2524-2530

[10] A. Ashikhmin, G. Kramer, S. ten Brink: "Code Rate and the Area under Extrinsic Information Transfer Curves", Proceedings of the *International Symposium on Information Theory*, 2002, p.115, IEEE

[11] The first part of this report: " On Extrinsic Information of Good Binary Codes Operating Over Gaussian Channels"